\documentclass[a4paper]{article}
\usepackage{latexsym}
\begin{document}

\makeatletter
\def\adots{\mathinner{\mkern 1mu\raise \p@ \vbox{\kern 7\p@ \hbox{.}}
\mkern 2mu \raise 4\p@ \hbox{.}\mkern 2mu\raise 7\p@ \hbox{.}\mkern 1mu}}
\makeatother

\newtheorem{theorem}{Theorem}
\newtheorem{proposition}{Proposition}
\newtheorem{remark}{Remark}
\newtheorem{corollary}{Corollary}
\newtheorem{lemma}{Lemma}
\newtheorem{observation}{Observation}

\newcommand{\qed}{\hfill$\Box$\medskip}

\newcommand{\vdashsp}{\vdash^{\,\,\:}}

\title{A SWAR Approach to Counting Ones} 
\author{Holger Petersen\\
Reinsburgstr. 75\\
70197 Stuttgart\\
Germany} 

\maketitle

\begin{abstract}
We investigate the complexity of algorithms counting ones in different sets of
operations. With addition and logical operations (but no shift)  $O(\log^2(n))$
steps suffice to count ones. Parity can be computed with complexity $O(\log(n))$,
which is the same bound as for methods using shift-operations. If multiplication 
is available, a solution of time complexity $O(\log^*(n))$ is possible improving 
the known bound $O(\log\log(n))$. 
\end{abstract}

\section{Introduction}  
The operation of counting the number of ones in a binary word consisting of $n$ bits is 
also known as sideways addition \cite{Knuth09}, bit count \cite{KR78}, 
or population count \cite{Chess}. Fast solutions based on look-up tables have been suggested by
several authors (see the references in \cite{Warren77}), but for large $n$ they are
clearly impractical.

Under unit cost measure, methods of asymptotical complexity $O(\log\log n)$
are known, but they are based on multiplication like the one due to Gillies and Miller from the first 
textbook on programming  published in 1957 \cite{WWG57,Knuth09} or 
division as in \cite[Item 169]{HAKMEM} (the algorithms are presented for 
specific word lengths and we assume a generalization in a natural way).
Since these more powerful arithmetical operations might not be executed as efficiently as,
e.g., addition or logical instructions, we are interested in
approaches to the bit count problem avoiding these operations, 
also ruling out shift instructions as special cases of multiplication and division.

Exercise~2-9 of \cite{KR78} asks for a solution of the bit count problem
based on an observation originally due to Wegner \cite{Wegner60}. It is a remarkable 
fact that in two's complement representation for $x\neq 0$ the right-most
one of $x$ can be deleted by the operation $x\,\&\, (x-1)$, where $\&$
denotes the bit-wise AND of two values. With the help of
this fact the bit count of the input can be computed in a loop that is
executed once for each one. This will
save considerable work in comparison to a naive implementation
when the number of ones is sparse. When the zeroes 
are sparse the method can be applied to the complement (this was already mention
in the last section of \cite{Wegner60}). The worst case as well as the average case 
complexity of these methods is however $\Theta(n)$ 
(for the average case consider a binary word $x$ with 
$\nu(x)$ ones and its complement having $n - \nu(x)$ ones).

In the current work we will develop algorithms for counting ones in a binary word
based on the concept of ``SIMD with a register (SWAR)'' \cite{FD98}. The idea is to
partition large registers into fields of bits that are processed in parallel.  We obtain
algorithms of complexity $O(\sqrt{n})$ and $O(\log^2n)$ in settings without multiplication 
and division, where the latter approach is less practical for current word-lengths as it requires a lot of
large constants.

Computing the parity of a binary word efficiently has received some
attention in its own right, see, e.g., \cite{Anderson,Brumme}. It can
of course be determined from the count of ones, but a specific approach might be faster.
For a modified parity function we get an 
$O(\log n)$ solution in a restricted computational model which is competitive with methods 
using shift instructions. 

If multiplication is included in the set of operations, the complexity of counting ones
can be reduced to $O(\log^*(n))$.
Also included is a parity function making use of multiplication and integer division 
which is superior (in terms of ``C-operations'') to the implementations the author is aware of.
It is inspired by the $O(\log^*(n))$ solution combined with 
\cite[Item 169]{HAKMEM}, which we outline in Appendix~B. 

\section{Preliminaries}

We will denote the binary logarithm by $\log$. The word-length will be denoted by
$n$, which is usually a power of $2$. Bit positions are numbered starting at $0$ with 
the least significant bit. Thus bits of the words being processed have weights $1$ to
$2^{n-1}$. Using the notation from \cite{Knuth09}, the bit count function is called
$\nu$.

The {\em tower function} is defined by $^02 = 1$ and 
$$^k2 = 2^{(^{(k-1)}2)} = 2^{2^{\adots^{2}}}$$
for $k > 0$ where $k$ is the number of twos.
The inverse of the tower function is the {\em iterated logarithm} $\log^*$. 
It can be extended to any $n\ge 1$ as the
number of times the  function $\log$ must be applied until the result
 is less than or equal to $1$

Our first results refer to programming models that rule out multiplication, division, 
and shift instructions.
The restricted set of instructions including only logical operations 
(and, or, xor) and addition which
will be called here Oblivious Parallel Addition and Logical Instructions (OPAL). Including
subtraction does not change this model in terms of asymptotic complexity, since it can be 
simulated with negation and addition. OPAL is the model employed in \cite{Warren77}.
Code will be presented in (subsets) of C \cite{KR78}.

With Parallel Addition and Logical Instructions (PAL) we will denote the extension by flow 
control instructions like ``if'' and ``while'', usually found in modern programming languages. 
Within the PAL model the bit count function from \cite{KR78} and the 
improvement of Exercise 2-9 can be realized directly.

\section{Results}

\begin{theorem}\label{palcountsqrt}
The bit count function $\nu$ can be computed with the help of  $O(\sqrt{n})$ instructions in the
PAL model.
\end{theorem}
{\bf Proof:} The central idea is to apply Wegner's technique \cite{Wegner60} to approximately 
$\sqrt{n}$ fields of  $\sqrt{n}$ bits each, separated
by spacer bits \cite{FD98}. Since the original input does not have spacer bits, the first task is to
count ones at the positions of the prospective spacer bits. This number is stored in the variable
{\tt sum}.

Then the following steps are repeated.
First the spacer bits are set. By subtracting twice the current most significant
bits of each field 
(plus one for the least significant
position), a one is deleted from all fields that contain a one. 
For fields without a one, the spacer bit
it reset by this operation. 
The difference to the previous state of the most significant bits
of the fields is computed and the change is determined with the help of the 
same technique. The current number of  ``active'' fields is stored in the variable {\tt count} and 
additional bits reset are
recorded in {\tt sum}. The subtraction from an empty field will cause a borrow and 
therefore no explicit
subtraction from the next field is necessary.
These steps are carried out until all fields are empty.

The number of iterations is bounded above by the length of the fields and thus 
$O(\sqrt{n})$. The amortized cost of updating variable {\tt count} is also $O(\sqrt{n})$,
since each of the $O(\sqrt{n})$ spacer bits can change only once
(it will never change from 0 to 1 within the loop). Therefore the claimed complexity
follows.
 
We present the procedure just outlined for an 32 bit input, 
where for the sake of simplicity we round 
the length and the number of fields to powers of 2:
\begin{verbatim}
#define HIBITSL 0x88888888  // 8 fields with 4 bits each
bitcount(a)  // ALU-based
register unsigned long a;
{
        register int sum;         // overall bit sum
        register int count;       // incremental contributions
        register unsigned long x; // auxiliary variable
        register unsigned long oldhi, newhi; // hi bits of fields

        x = (a & HIBITSL);
        for(sum = 0; x ; x &= (x-1))
                sum++;        
        count = 8;   // bitcount(HIBITSL)
        oldhi = HIBITSL;
        a |= oldhi;
        while(oldhi)
                {
                        a &= (a - oldhi - oldhi - 1);
                        newhi = (a & oldhi);
                        x = newhi ^ oldhi;
                        while(x) {
                                x &= (x-1);
                                count--;
                        }
                        oldhi = newhi;
                        sum += count;
                };
        return(sum);
}
\end{verbatim}
\qed

Next we exclude flow control ({\tt if}, {\tt while}) from the operations allowed. By the result
from \cite{Warren77}, function $\nu$ cannot be computed under this restriction. We can however
compute a modified version of $\nu$ for a portion of the bits.

First we describe an important building block that allows us to shift bits rapidly using
only addition and logical operations.

\begin{lemma}\label{shift}
In a set of disjoint fields, the least significant bit of each field can be shifted to the most
significant position in constant time, if all intermediate positions contain zero.
\end{lemma}
{\bf Proof:} We will describe the technique for a single field stretching from 
position  $i$ to position $j > i$.  By incorporating information for the other fields into
the constants, the corresponding modifications can be achieved.

First bit $j$ is cleared by masking with $\neg 2^j$.
Then the value $2^j-2^i$ is added to the result. Finally we mask with
$\neg(2^j-2^i)$.

If bit $i$ initially was $1$, then there will be a carry that propagates
to position $j$. If it was $0$, then position $j$ will not be modified. \qed  

We note that the technique of Lemma~\ref{shift} can replace the
bit-by-bit shift of bits in the construction from \cite{Warren77}, which
has a time complexity proportional to the number of positions.

\begin{theorem}\label{opalcount}
Let $m = n - \lceil\log n\rceil$.
The modified bit count function $2^{m-1}\nu(x\bmod 2^m)$ for the lowest $m$ bits 
of an $n$ bit input $x$ can be computed with the help of  $O(\log^2(n))$ instructions in the
OPAL model.
\end{theorem}
{\bf Proof:} We will demonstrate how to compute the function for the lower $n/2$ bits of a word with
$n$ bits. The claim then follows by computing in addition the function for the most significant
$n/2 - \lceil\log n\rceil$ bits (shift all constants accordingly), adding the results and subtracting the
count of
ones in the overlap by first shifting each bit to the target area and then subtracting it. 
The latter 
correction can be done in time $O(\log n)$ by Lemma~\ref{shift}.

In stage $i$ of the method, counts of ones in fields of length $2^{i-1}$ are combined into
counts for fields of length $2^{i}$ by applying Lemma~\ref{shift} the bits of the counts. 
Since by Lemma~\ref{shift} we can only shift left, the counts are are located at
the most significant bit of fields.

There are $O(\log n)$ stages and each stage takes time $O(\log n)$, resulting in the claimed
bound $O\log^2(n)$

The following code implements the method for the lower 16 bits of a 32 bit word.
Binary representations of the masks are given in the comments
\begin{verbatim}
// least significant 16 bits counted, 20 bits required 
// (excess 4 bits for 5 bit count)
bitcount(x)
register unsigned  x;
{
        register unsigned y;

        y = x & 0x5555;   // 0101010101010101   
        x -= y;           // delete fields
        y +=  0x5555;     // 0101010101010101   
        y &= 0xAAAA;      // 1010101010101010   
        x += y;

        y = x & 0x6666;   // 0000110011001100110
        x -= y;           // delete fields
        y += 0xCCCC;      // 0001100110011001100 
        y &= 0x73333;     // 1110011001100110011
        y += 0x6666;      // 0000110011001100110
        y &= 0x79999;     // 1111001100110011001
        x += y;

        y = x & 0x3838;   // 0000011100000111000
        x -= y;           // delete fields
        y += 0x1E1E0;     // 0011110000111100000
        y &= 0x61E1F;     // 1100001111000011111
        y += 0xF0F0;      // 0001111000011110000   
        y &= 0x70F0F;     // 1110000111100001111
        y += 0x7878;      // 0000111100001111000  
        y &= 0x78787;     // 1111000011110000111
        x += y;

        y = x & 0x780;    // 0000000000000011110000000
        x -= y;           // delete fields
        y += 0x3FC00;     // 0000000111111110000000000 
        y &= 0x7C03FF;    // 0011111000000001111111111
        y += 0x1FE00;     // 0000000011111111000000000
        y &= 0x3E01FF ;   // 0001111100000000111111111
        y += 0xFF00;      // 0000000001111111100000000 
        y &= 0x1FF00FF;   // 1111111110000000011111111
        y += 0x7F80;      // 0000000000111111110000000
        y &= 0xF807F;     // 0000011111000000001111111 
        x += y;

        return(x);
}
\end{verbatim}
\qed 

By some pre- and post-processing we can obtain a solution in the PAL model 
of the same asymptotic complexity. Due to the large constant factor it does however not appear to be of 
practical value for small word-lengths.

\begin{corollary}\label{palcountlog}
The bit count function $\nu(x)$ 
of an $n$ bit input $x$ can be computed with the help of  $O(\log^2(n))$ instructions in the
PAL model.
\end{corollary}
{\bf Proof:} We proceed as follows:
\begin{enumerate}
\item\label{step1}
The $\Theta(\log n)$ most significant positions not handled by the method of 
Theorem~\ref{opalcount} are copied to a variable and set to 0 in $x$ 
(time complexity $O(1)$). 
\item\label{step2}
For the $n - \Theta(\log n)$ bits the modified bit count function is computed according to
Theorem~\ref{opalcount} (time complexity $O(\log^2 n)$). 
\item\label{step3}
The resulting $O(\log n)$ bits are transferred one by one to the 
least significant positions by extracting each bit with the help of a mask and building
up the result in another variable (time complexity $O(\log n)$).
\item
The most significant bits of the original input 
saved in step~\ref{step1}) are counted in a naive way 
and added to the result of step~\ref{step3}) (time complexity $O(\log n)$).
\end{enumerate}
The overall time complexity is dominated by step~\ref{step2}) and thus $O(\log^2 n)$.
\qed

For the parity function only a single bit has to be shifted in an approach 
as in Theorem~\ref{opalcount}. Therefore we obtain a more efficient solution than for
bit count.
\begin{theorem}
The modified parity function $2^{n-1}(\nu(x) \bmod 2)$ can be computed with the help of  $O(\log n)$ instructions in the
OPAL model.
\end{theorem}
{\bf Proof:} In the initial stage the bits of the input are moved up one position and the parity of  pairs of bits is computed by a XOR. 
By masking out the lower bits of each pair, we obtain 2 bit fields containing the parity in the most significant bit.

In the following stages we apply a variant of  Lemma~\ref{shift}, where we propagate the most significant bit of one field directly to
the most significant bit of the next field. We make use of the fact that in the least significant positions of each field spacer bits
in the sense of \cite{FD98} are available after masking the most significant bit. Each stage doubles the size of the fields.

We illustrate the method for 32 bits in the following code:  
\begin{verbatim}
parity(x)
register unsigned x;
{
        x ^= (x + x); 
        x &= 0xaaaaaaaa;  // parity of 2 bit fields
        x += 0x66666666;   
        x &= 0x88888888;  // 4 bit fields
        x += 0x78787878;
        x &= 0x80808080;  // 8 bit fields
        x += 0x7f807f80;
        x &= 0x80008000;  // 16 bit fields
        x += 0x7fff8000;
        x &= 0x80000000;  // all 32 bits
        return(x);
}
\end{verbatim}
\qed

By the result from \cite{Warren77} the OPAL model is not able to 
directly compute the parity function. With the help
of a single test a non-zero result can however be transformed into $1$. We thus obtain:
\begin{corollary}
The parity function $\nu(x) \bmod 2$ can be computed with the help of  $O(\log(n))$ instructions in the
PAL model.
\end{corollary}

We now turn to a more powerful model of computation.
\begin{theorem}\label{mulcount}
The bit count function $\nu$ can be computed in time $\log^*(n)$ with the help of addition, 
shift, logical operations, and multiplication. 
\end{theorem}
{\bf Proof:} The algorithm proceeds in iterations. 
We will maintain the invariant that at the end of iteration $k$ of the algorithm
fields of length $^k2$ of the current intermediate
result $a$ contain a count of ones in corresponding positions of the original input. 
The count will 
be stored in the low order bits of each field.
Initially the invariant holds for $k=0$ and the input $a$, since a single bit represents its own count.

Consider $k > 0$ and assume the invariant holds for the current intermediate
result $a$ after iteration $k-1$. We define bit mask
$$ b_k = \cdots \underbrace{11\cdots 11}_{^{(k-1)}2}\underbrace{00\cdots 00}_{^{(k-1)}2}\underbrace{11\cdots 11}_{^{(k-1)}2}$$
that selects half of the fields from the previous iteration and 
$$ c_k= \cdots\underbrace{00\cdots 00}_{^k2-2\cdot^{(k-1)}2}\underbrace{11\cdots 11}_{2\cdot^{(k-1)}2}\underbrace{00\cdots 00}_{^k2-2\cdot^{(k-1)}2}\underbrace{11\cdots 11}_{2\cdot^{(k-1)}2}$$
that selects the lower $2\cdot^{(k-1)}2$ bits of each field of length $^k2$. 
We let $x = a \& b_k$, $y = (a >> ^{(k-1)}2)  \& b_k$. Next we multiply $x$ and $y$ with 
a number 
$$ m_k = \underbrace{00\cdots 01}_{2\cdot ^{(k-1}2}\underbrace{00\cdots 01}_{2\cdot ^{(k-1)}2}\underbrace{00\cdots 01}_{2\cdot ^{(k-1)}2}$$
that contains $^k2/(2\cdot ^{(k-1)}2)$ ones and form 
$$a = ((x \cdot m_k)  >> (^k2-^{(k-1)}2) \& c_k) + ((y \cdot m_k)  >> (^k2-^{(k-1)}2) \& c_k).$$ 
Notice that in each field of length $2\cdot^{(k-1)}2$ at most $^k2$ values of size
less than $2^{(^{(k-1)}2)} = ^k2$ are added. Therefore the sum is less than $(^k2)^2$ and fits into
$2\cdot^{(k-1)}2$ bits.
The process stops when $^k2 = 2^{(^{(k-1)}2)} > n$ at the start of iteration $k$ 
and a field of $^{(k-1)}2$ bits can store the count of all bits of the input in binary. 

Then we multiply the intermediate result with
$$ \underbrace{00\cdots 01}_{^{(k-1)}2}\underbrace{00\cdots 01}_{^{(k-1)}2}\underbrace{00\cdots 01}_{^{(k-1)}2}$$
containing $n/(^{(k-1)}2)$ ones and shift right by $n - ^{(k-1)}2$ positions to obtain the result.
\qed 

We finally include code for computing the parity function by combining ideas of
the bit count function from Theorem~\ref{mulcount} with an initial stage of an
 $O(\log(n))$ method. By reducing the number of possible ones via a XOR,
the fields can be smaller than in the more general setting of a full count.
The final multiply and shift of  Theorem~\ref{mulcount} is replaced by one division as in
\cite[Item 169]{HAKMEM} to obtain the following solution, which uses
only 7 ``C-operations'' as opposed to the 8 operations of the ``parity of word with a multiply''
from \cite{Anderson}.

\begin{samepage}  
\begin{verbatim}
parity(x)
register unsigned x;
{
        x ^= x >> 1;      // parity of 2 bits
        x &= 0x55555555;  // select low bits of 2 bit fields
        x *= 0x15;        // add three 2 bit fields
        x &= 0x41041041;  // select low bits of 6 bit fields        
        return((x % 0x3f) & 0x1); // apply HAKMEM technique,
                                  // and return least sign. bit
}
\end{verbatim}
\end{samepage} 

This approach would work for word-lengths up to 372 (with constants adjusted).

\section{Discussion}

We have obtained several bit count and parity algorithms for sets of operations not including
multiplication or division.
Asymptotically the (modified) parity function with complexity $O(\log n)$
for the OPAL-model is as fast as the 
known solutions based on broadword steps \cite{Knuth09}. The latter may include shift-operations
by a constant number of bit positions. 

In a setting with the powerful arithmetical instruction multiplication
we could improve the complexity of counting ones
from  $O(\log\log(n))$ to $O(\log^*(n))$.

The following table summarizes upper time bounds for counting ones with different sets of operations.
\begin{center}
\begin{tabular}{|l|c|c|}\hline
Operations                 & depending on $n$ & dep.\ on $\nu(x)$  \\\hline\hline
increment, decrement, & $O(n)$  &  $O(\nu(x))$\\
 logical operations & see Appendix A &  \cite{Wegner60} \\\hline
addition, logical operations  & $O(\log^2(n))$ & \\
(PAL)  &   Cor.~\ref{palcountlog} &   \\\hline
addition, shift,  logical operations  & $O(\log(n))$& \\
(broadword steps) &  folklore, see \cite{Chess}  &   \\\hline
addition, shift, logical operations, & $O(\log^*(n))$  & $O(\log\log(\nu(x)))$ \\
multiplication  &  Thm.~\ref{mulcount} & \cite{Petersen15a}   \\\hline
addition, shift, logical operations, & $O(\log\log(n))$  &  \\
division  &  \cite[Item 169]{HAKMEM}  &   \\\hline
\end{tabular}
\end{center}
For most sets of operations we are not aware of algorithms depending on 
$\nu(x)$ and thus taking advantage of sparse inputs.

A lower time bound $\Omega(\log n/\log\log n)$ on parity (and thus counting ones) 
in the model of broadword steps follows from a result in circuit complexity 
(see \cite[Ex-~127]{Knuth09}). Although increment, decrement, and 
logical operations seem to admit no efficient algorithms, we were not able to prove
a lower bound $\Omega(\nu(x))$ corresponding to the upper bound from \cite{Wegner60}.

\section*{Appendix A}
The algorithm from \cite{Wegner60} of complexity $O(\nu(x))$ is based
on increment, decrement, and logical operations. In contrast the standard
method of complexity $O(n)$
presented in \cite{Wegner60} as well as the approaches in \cite{Frieden60}
and \cite[p.~47]{KR78} require shift- or rotate-operations. We therefore
include an algorithm of complexity $O(n)$ that uses the more restricted
set without shift/rotate and general addition. 
It is based on technique (2) from \cite{Warren77}.
\begin{verbatim}
bitcount(a)
register unsigned long a;
{
        register int sum;         //  bit count
        register unsigned long mask; // bit mask

        mask = 1;
        sum = 0;
        while(mask)
                {
                        if (a & mask) sum++ ;
                        mask |= (mask-1); // propagate rightmost 1
                        mask++;         // new bit mask
                };
        return(sum);
}
\end{verbatim}

\section*{Appendix B}
HAKMEM item 169 \cite{HAKMEM} describes a reduction of bit counting for word-lengths of at most 62 bits to 
integer division. In its original form the program is presented in
assembly language for the PDP-6/10 (36 bit architectures), which is
not very well-known nowadays.%
\footnote{Most of the PDP-6/10 instructions are quite suggestive. Two possible exceptions
are:
\begin{description}
\item[{\tt LDB B,[014300,,A]}:] Load 35 bits (octal {\tt 043}) from 
{\tt A} with an offset 1, counting bits from the right. Store them right adjusted in 
{\tt B} (line 1 of the C program, 
this cannot be implemented directly in C and we use the approach mentioned 
in the comment of  \cite[item 169]{HAKMEM} on the {\tt LDB} instruction).
\item[{\tt SUBB A,B}:] Subtract and store in both {\tt A} and {\tt B} (lines 6 and 7 of the C program).
\end{description}
See \cite{pdp10} for more information on the PDP-6/10 instruction set.}

We render it here in C (where the variable names {\tt a} and 
{\tt b} correspond to the registers of the original program,  comments are given separately):

\begin{verbatim}
#define TWOBITS 033333333333
#define THREEBITS 030707070707

bitcount(a)
unsigned a;
{
        unsigned b;
     
        b = a>>1;        // line 1
        b &= TWOBITS;    // line 2
        a -= b;          // line 3
        b >>= 1;
        b &= TWOBITS;
        a -= b;          // line 6
        b = a;           // line 7
        b >>= 3;
        a += b;          // line 9
        a &= THREEBITS;
        return(a % 077); // line 11
}
\end{verbatim}
Comments:
\begin{description}
\item[{\tt lines 1-6}]: Consider an octal digit of variable {\tt a} 
 composed of three bits $x$, $y$, and $z$
 with weight $4x + 2y + z$. Then in line 3 the value $2x + y + z$ is computed and in line 6
 the sum $x + y + z$ of the three original bits is computed.
\item[{\tt line 7}]: This simulates the second transfer of {\tt SUBB}.
\item[{\tt line 9}]: Neighboring groups of octal digits are added in order to compute 
 the sum of six bits. Notice that the three resulting bits are able to hold the maximum count
 for 6 bits without a carry to the next octal digit. 
\item[{\tt line 11}]: Consider the contents of {\tt a} as a number in base 64 representation. Then
 the digit at position $i$ from the right (starting at 0) has weight 
$$64^i = (63 + 1)^i = 63^i + 63^{i-1}i + \cdots + 63i + 1$$
amd modulo $63$ the sum of all base $64$ digits is computed. This limits the admissable 
word-length to 62 bits. 
\end{description}



\end{document}